\documentclass{sig-alternate-10pt}
\usepackage{url}
\usepackage{graphicx}
\usepackage{cite}
\usepackage{tabularx}
\usepackage{array}
\usepackage{color}
\usepackage{arydshln}
\usepackage{multirow}
\usepackage[labelfont=bf]{caption} 

\newcommand{\squeezeup}[1]{\vspace{-#1mm}}
\newcommand{\pushup}[1]{\vspace{#1mm}}

\newcommand{\colspace}[1]{\setlength\tabcolsep{#1mm}}
\newcolumntype{C}[1]{>{\centering\let\newline\\\arraybackslash\hspace{0pt}}m{#1}}
\newcommand{\captionnote}{\fontsize{7.2}{7.5}\selectfont}
\usepackage{amsmath}
\usepackage{algorithm}
\usepackage[noend]{algpseudocode}

\makeatletter
\def\BState{\State\hskip-\ALG@thistlm}
\makeatother

\begin{document}
%\mainmatter
\title{Towards Characterizing International Routing Detours}

\author{
    Anant Shah\\
   %     \normalsize{Computer Science Dept.}\\
        \normalsize{Colorado State University}\\
        \texttt{akshah@cs.colostate.edu}
\and
Romain Fontugne\\
      %  \normalsize{Internet Initiative Japan (IIJ)}\\ 
        \normalsize{IIJ Research Lab}\\
        \texttt{romain@iij.ad.jp} 
  \and
    Christos Papadopoulos\\
  %      \normalsize{Computer Science Dept.}\\
        \normalsize{Colorado State University}\\
        \texttt{christos@cs.colostate.edu}
}
\maketitle

% !TEX root = main.tex
\begin{abstract}

There are currently no requirements (technical or otherwise) that BGP paths must be contained within national boundaries.  Indeed, some paths experience \emph{international detours}, i.e., originate in one country, cross international boundaries and return to the same country.  In most cases these are sensible traffic engineering or peering decisions at ISPs that serve multiple countries.  In some cases such detours may be suspicious. Characterizing international detours is useful to a number of players: (a) network engineers trying to diagnose persistent problems, (b) policy makers aiming at adhering to certain national communication policies, (c) entrepreneurs looking for opportunities to deploy new networks, or (d) privacy-conscious states trying to minimize the amount of internal communication traversing different jurisdictions.

In this paper we characterize international detours in the Internet during the month of January 2016. To detect detours we sample BGP RIBs every 8 hours from 461 RouteViews and RIPE RIS peers spanning 30 countries. Then geolocate visible ASes by geolocating each BGP prefix announced by each AS, mapping its presence at IXPs and geolocation infrastructure IPs. Finally, analyze each global BGP RIB entry looking for detours. Our analysis shows more than 5K unique BGP prefixes experienced a detour. A few ASes cause most detours and a small fraction of prefixes were affected the most. We observe about 544K detours. Detours either last for a few days or persist the entire month. Out of all the detours, more than 90\% were \emph{transient} detours that lasted for 72 hours or less. We also show different countries experience different characteristics of detours. 
\end{abstract}

\squeezeup{1}
% !TEX root = main.tex
\section{Introduction}
\label{sec:intro}
We define an international detour (detour for short) as a BGP path that originates in an AS located in one country, traverses an AS located in a different country and returns to an AS in the original country. Detours have been observed in the Internet, for example, cities located in the African continent communicating via an external exchange point in Europe\cite{PAM14:Africa}. Many autonomous systems are also multinational, which means that routes traversing the AS may cross international boundaries. There have also been suspicious cases of detours. In November, 2013, the Internet intelligence company Renesys (now owned by Dyn) published an online article detailing an attack they called Targeted Internet Traffic Misdirection~\cite{COWIE:2013:Online}. Using \texttt{Traceroute} data they discovered three paths that suffered a man-in-the-middle (MITM) attack. One path originated from and was destined to organizations in Denver, CO, after passing through Iceland, prompting concern and uncomfortable discussions with ISP customers.
Each of these anecdotes, while interesting in its own right, does not address the broader question about how prevalent such detours are, their dynamics and impact. Characterizing detours is important to several players: (a) as a tool for network engineers trying to diagnose problems; (b) policy makers aiming at adhering to potential national communication policies mandating that all intra-country communication be confined within national boundaries, (c) entrepreneurs looking for opportunities to deploy new infrastructure in sparsely covered geographical areas such as Africa, or (d) privacy-conscious states trying to minimize the amount of internal communication traversing different jurisdictions. Using the methodology developed to detect detours we also present a tool, \texttt{Netra}\footnote{\url{https://github.com/akshah/netra}}, to monitor the Internet routing system in near real-time and produce alerts. Network operators can not only appear informed about the incident, but also may be able to take action in peer selection in response to the alerts. Finally, longitudinal analysis of detours can give us insight into how routing and network infrastructure evolve over time. 

%Previous research has looked at a similar problem and the two most closest to ours are~\cite{PAM14:Africa} and~\cite{boomerang}. Both examine detours from the data plane point of view, while we use control plane information. In~\cite{PAM14:Africa} researchers explore large latency paths in Africa and discover that they often detour through Europe. Closer to our work is~\cite{boomerang}, but at a much smaller scope. We present other related work in Section~\ref{sec:related}. 
In this paper we first develop methodology to detect detours, validate it on live traffic using our tool \texttt{Netra} and then use it to characterize them at a global scale on historical BGP data of January 2016 from RouteViews and RIPE RIS. 
%The reason we study only one month is that this is an initial study of detours to determine their scope and magnitude, as well as develop a methodology for analysis. Moreover, the computation required is significant, taking a few weeks to analyze a month's worth of data.

The rest of this paper is organized as follows. In Section~\ref{sec:related} we present related work and highlight previous efforts in direction similar to ours and point out key areas where our work differs from them. In Section~\ref{sec:datasources} we describe our datasets, corresponding usage and reasoning for the choice of our datasets. Section~\ref{sec:asgeolocation} details the methodology used to perform AS geolocation and analysis of our geolocation results. Section~\ref{sec:pathanalysis} explains in detail detour detection process, corresponding terminologies used throughout the paper. In Section~\ref{sec:detourvalidation} we explain our data plane measurements and present validation results. In Section~\ref{sec:results} we characterize detours seen in January 2016. First we present aggregate analysis of entire dataset, then classify detours into different categories and finally focus on \emph{transient detours} in Sections~\ref{sec:aggresults}, \ref{sec:detoursclassification} and \ref{sec:characterizingtransientdetours} respectively. In Sections~\ref{sec:discussion} and \ref{sec:conclusion} we discuss value additions of our work, summarize and present future work respectively.
\squeezeup{1}
% !TEX root = main.tex
\section{Related Work}
\label{sec:related}
\noindent
\textbf{Detour detection:}\\
In November 2013 Renesys reported a few suspicious paths~\cite{COWIE:2013:Online}. One went from Guadalajara, Mexico to Washington, D.C. via Belarus; another went from Denver, CO through Reykjavik, Iceland, back to Denver. They used mostly data plane information from traceroute for their analysis. In \cite{PAM14:Africa} the authors focus on ISP inter-connectivity in the continent of Africa. They searched for paths that leave Africa only to return back. The goal, however, was to investigate large latencies in Africa and ways to reduce it. The premise was that if a route crosses international boundaries it would exhibit high latency. The work pointed to cases where local ISPs are not present at regional IXPs and IXP participants don't peer with each other. Similar to Renesys, they also use traceroute measurements, this time from the BISmark infrastructure (a deployment of home routers with custom firmware) in South Africa. Our study extends beyond Africa and investigates transient in addition to long-lasting detours. In \emph{Boomerang}~\cite{boomerang}, the authors again use traceroute to identify routes from Canada to Canada that detour through the US. In this work the motivation was concerns about potential surveillance by the NSA. This work differs from ours in a number of ways: we characterize detours not just for one but 30 countries using control plane information rather than data plane. We use data plane measurement only for validation purposes. Our goal is to not only detect detours but show characteristics about them which previous work does not present. \\\\
\noindent
\textbf{Data plane vs Control plane Incongruities:}\\
In~\cite{routingPolicies} authors focus on routing policies and point out cases where routing decisions taken by ASes do not conform to expected behavior. There are complex AS relationships, such as, hybrid or partial transit which impact routing. Such relationships may lead to false positives in our results. However, the paper points out that most violations of expected routing behavior caused by complex AS relationships are very few and most violations were caused by major content providers. Our work identifies detours for variety of ASes, including both large content providers and small institutions. Moreover, in \cite{quantifyingpitfalls} authors argue that such incongruities are caused due to incorrect IP to AS mappings. About 60\% of mismatches occur due to IP sharing between adjacent ASes. Authors here show that 63\% to 88\% of paths observed in control plane are valid in data plane as well. The work in~\cite{wiretapping} also analyzes the control plane (RIBs and AS paths) to construct a network topology and then uses traceroute to construct country-level paths. The goal of this work was to understand the role of different countries that act as hubs in cross-country Internet paths. Their results show that western countries are important players in country level internet connectivity.\\\\
\noindent
\textbf{Malicious AS detection:}\\
In~\cite{ASwatch} authors present \emph{ASwatch}, an AS reputation system to detect bulletproof hosting ASes. Similar to our work ASwatch relies on control plane information to detect malicious ASes (that may host botnet C\&C servers, phishing sites, etc). The motivation of this work is different than ours. ASwatch attempts to detect malicious ASes by mining their link stability, IP space fragmentation and prefix reachability. ASwatch will not detect ASes that cause detours. The detour origin ASes that our work detects could complement features that ASwatch uses. As authors in \cite{ASwatch} point out malicious ASes rewire their routes more frequently than legitimate ones, transient detours might be particularly useful to improve detection capability of ASwatch.\\\\
\noindent
\textbf{Geolocation Accuracy:}\\
In context of MaxMind geolocation accuracy, \cite{geolocationcompare} and \cite{studygeolocation} have shown MaxMind country geolocation to be 99.8\% in consensus with other geolocation DBs. In~\cite{germanyPAM2012} authors use data from Routing Information Registries (RIRs), RIPE DB and Team Cymru to determine all IP blocks and ASes that geolocate to Germany. To validate their geolocation accuracy, authors query the MaxMind database which allows mapping IP addresses to their country of presence. We adopt a more exhaustive strategy than~\cite{germanyPAM2012}.\\\\
\noindent
\textbf{Control-plane-only for detection:}\\
One way to detect detours is to use \emph{traceroute}, analyze reported hops and use latency as an indication of a detour. This approach was followed by~\cite{PAM14:Africa} that studied peering relationships in Africa; we too use this approach to validate our results on live data. However, we detect detours using only control plane data. This has a number of advantages: 1) Collecting data plane information at an Internet scale is hard. It needs infrastructure and visibility provided by Atlas probes or Ark monitors is limited. Moreover, running too many traceroutes from own network to others might lead to blacklisting. 2) Small footprint of our methodology makes it easily reproducible. Any network operator can pull a RIB dump from his/her border router and run \texttt{Netra} to detect detours for prefixes they own.

\squeezeup{1}
% !TEX root = main.tex
\section{Data Sources}
\label{sec:datasources}

\begin{table*}[htb!]
\colspace{2}
\centering
\small
\caption{Dataset Description}
\begin{tabular}{C{2.5cm}:C{2.8cm}:C{2.5cm}:C{3cm}:C{3.3cm}}
\hline
 Name & Usage & Date & Sources & Info\\ \hline
BGP & AS Geolocation; Detour Detection & 2016-01 &RouteViews, RIPE RIS & 38,688 RIBS, 416 peers, 30 countries, 55GB \\\hdashline
Infrastructure IP List & AS Geolocation & 2016-01 to 2016-03 &CAIDA Ark, iPlane, OpenIPMap, RIPE Atlas Measurements & 3M Router IPs\\\hdashline
Infrastructure IPs to AS Mapping & Infrastructure IP geolocation & 2015-08 & CAIDA ITDK, iPlane& 6.6M IP to AS mappings\\\hdashline
AS to IXP Mapping & AS Geolocation & 2016-01 to 2016-03 &IXP websites, PeeringDB, PCH & 368 IXP websites crawled\\\hdashline
AS Relationship & Filtering peered paths from detection & 2016-01 & CAIDA AS Relationship & 482,657 distinct relationships\\\hdashline
Traceroute & Detour Validation & 2016-05-01 & RIPE Atlas & Used by \texttt{Netra}, 163 traceroutes \\\hdashline
MaxMind & Prefix Geolocation; Detour Validation & 2016-01, 2016-03 &MaxMind GeoLite City (free and paid) & Paid version used only for geolocating infrastructure IPs and detour validation\\\hline
\end{tabular}
\label{table:datasetinfo}
\end{table*}

We use variety of data sources to perform AS geolocation, BGP RIBS for detour detection and Traceroutes from RIPE Atlas for detour validation. In Table~\ref{table:datasetinfo} we list different datasets with their usage and relevant information about each. Our sampling rate is 3 RIBs per day (one every eight hours, as provided by RIPE RIS) for a total of 38,688 RIBs from 416 peers. This spans 30 countries, which amounts to about 55GB of compressed MRT data. We acknowledge that 30 countries do not necessarily represent global scale, but our scope is limited by placement of peers that provide BGP feeds. We used all v4 peers in our analysis. 

For geolocation of IP addresses we use MaxMind GeoLite City DB\cite{MaxMind}. We treat end user IPs and infrastructure IPs differently since MaxMind is known to be more accurate for eye-ball networks only. To gather the list of infrastructure IPs we used list of routers from CAIDA Ark traceroutes\cite{caida:ark}, OpenIPMap\cite{openipmap}, iPlane\cite{iplane} and RIPE Atlas built-in measurements and the anchoring measurements. The built-in measurements use all the RIPE Atlas probes and the destinations are root servers. The anchoring measurements are from 400 Atlas probes to other 189 Atlas anchors. These infrastructure IPs are then mapped to AS using IP to AS mappings from CAIDA ITDK\cite{ITDK:Online}, iPlane or longest prefix match. 

In addition to BGP sources, we use AS-to-IXP mapping to estimate presence of an AS in a country. We gather AS to IXP mappings from Packet Clearing House (PCH)\cite{pch}, PeeringDB\cite{peeringdb} and by crawling 368 IXP websites that make their participant list public. Finally, we use CAIDA AS Relationship datasets\cite{caida:asrel} to eliminate false positives from detours detected. In Section~\ref{sec:asgeolocation} we provide more details on how these datasets are used in AS geolocation along with a flowchart (Figure~\ref{fig:asgeoflowchart}).\\
\squeezeup{1}
% !TEX root = main.tex

\section{AS Geolocation}
\label{sec:asgeolocation}
 
To detect detours we are interested in country level geolocation. We define AS geolocation as presence of an AS in a country. An AS can have presence in multiple countries, especially ASes that belong to large providers. We detect the presence of an AS in country \emph{A} if it :
\begin{enumerate}
\item{Announces a prefix that geolocates to \emph{A} or}
\item{Has infrastructure IPs that geolocate to \emph{A} or}
\item{Has a presence at an IXP in \emph{A}}.
\end{enumerate}

\begin{figure*}[htb]
\centering
	\includegraphics[scale=0.5]{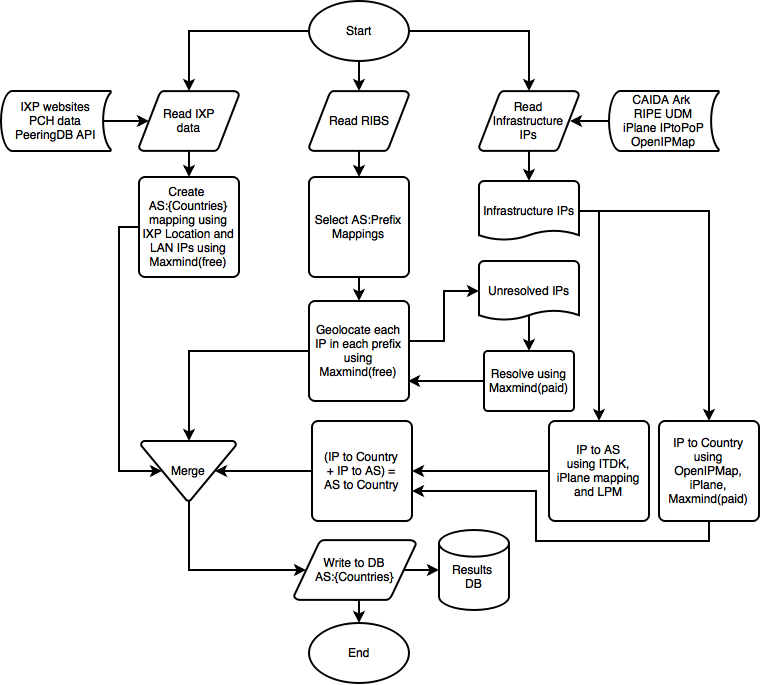}
	\caption{Flowchart: AS-to-Country mapping creation}
	\label{fig:asgeoflowchart}
\end{figure*}

In Figure~\ref{fig:asgeoflowchart} we show a flowchart detailing AS geolocation processes. There are 3 main steps as described above. In next sections we elaborate on each.

%Methodology flowchart
\subsection{Prefix Geolocation}
\label{sec:prefixgeolocation} 
We begin by geolocating all advertised BGP prefixes by an AS. It is possible that during our analysis in January 2016 some AS erroneously announced prefixes that it did not own. Therefore we perform a simple filtering; we trust an AS to be owner of a prefix if it announced the prefix for at least 15 days in our dataset. We assume most mistakes or hijacks will be less than this duration. Next, to map a BGP prefix to a country we geolocate each IP in the prefix using MaxMind-free. MaxMind could not geolocate 3.8M IPs. We could successfully geolocate 1.48M of these IPs with MaxMind-paid, we could not use remaining 2.32M IPs. Now we use the union of IP geolocation sets to get the BGP prefix geolocation. Due to the 2.32M IPs not geolocating even with paid version of MaxMind, 614 BGP prefixes could not be geolocated. For the remaining 610,722 BGP prefixes\footnote{\small{We use BGP prefixes `as is' from the RIBs and do not perform any prefix aggregation. For example, if both /8 and /9 blocks of a prefix were seen in RIBs of the same or different peers, they are treated as 2 separate prefixes.}} which were geolocated we observe that more than 99\% geolocated to single country. We note that 328,398 BGP prefixes were /24s. When BGP prefixes map to more than one country, the average size of the set was 2.9 countries. Finally, we perform union of geolocation sets of all BGP prefixes that an AS announces to create $1^{st}$ AS to country set. 

%%Prefix Geolocation Example
%\begin{figure}
%\centering
%	\includegraphics[scale=0.5]{figures-2014/methodologyprefix.png}
%	\caption{Example showing geolocation of a prefix seen in RIB}
%	\label{fig:prefixgeolocationexample}
%%\end{minipage}
%\end{figure}
\subsection{Infrastructure IP Geolocation}
As mentioned previously, we treat infrastructure IP addresses separately. Router geolocation is known to be inaccurate~\cite{routergeolocation}. Therefore for these IPs we want to create country geolocation set as large as possible. We populate list of router IPs from CAIDA Ark Traceroutes, iPlane IP to PoP mappings, OpenIPMap and RIPE built in measurements. Our list included 3M router IPs. This is the `Read Infrastructure IPs' step shown in flowchart Figure~\ref{fig:asgeoflowchart}. To geolocate each router IP we look at country location provided by iPlane, OpenIPMap\footnote{\small{OpenIPMap is crowdsourced and may not be very accurate. We use cases where confidence level for router geolocation is higher than 90\%.}} and Maxmind-paid and perform a union to give a set of countries. Next step is to map these routers to ASes. IP to AS is a challenging problem and active area of research. We use the best datasets available to create these mappings. Both CAIDA ITDK and iPlane datasets provide IP to AS mappings using the methodology described in~\cite{topologydualism}. For cases where either of these datasets fail to provide IP to AS mapping, we perform longest prefix match on the global routing table and map the IP to the AS announcing the longest matching prefix. Lastly, we combine IP to Country and IP to AS mappings to give $2^{nd}$ AS to country set. 

\subsection{IXP Presence of an AS}
\label{sec:ixppresence}
We extract presence of ASes at different IXPs and add the geolocation of IXP to the AS geolocation. As shown in `Read IXP data' step in Figure~\ref{fig:asgeoflowchart}, we use 3 sources of AS to IXP mappings. First, we crawl 368 IXP websites and extract their corresponding participants. Next, we use PeeringDB 2.0 API \cite{peeringdb} and lastly, we use dataset from Packet Clearing House (PCH) that lists participants at IXPs that PCH is also a part of.  We then combine geolocation obtained from these IXP sources to obtain $3^{rd}$ AS to country set. We acknowledge that IXP mappings from websites, PCH and PeeringDB might not be updated regularly and hence lead to mapping of an AS to a country that it does not have a presence in. Note that this will lead to false negative (not false positives) in detour detection, a trade-off we make to error on safe side. 

\subsection{AS to Country Set}
\begin{figure}[htb]
\centering
	\includegraphics[width=2.5in,height=1.6in]{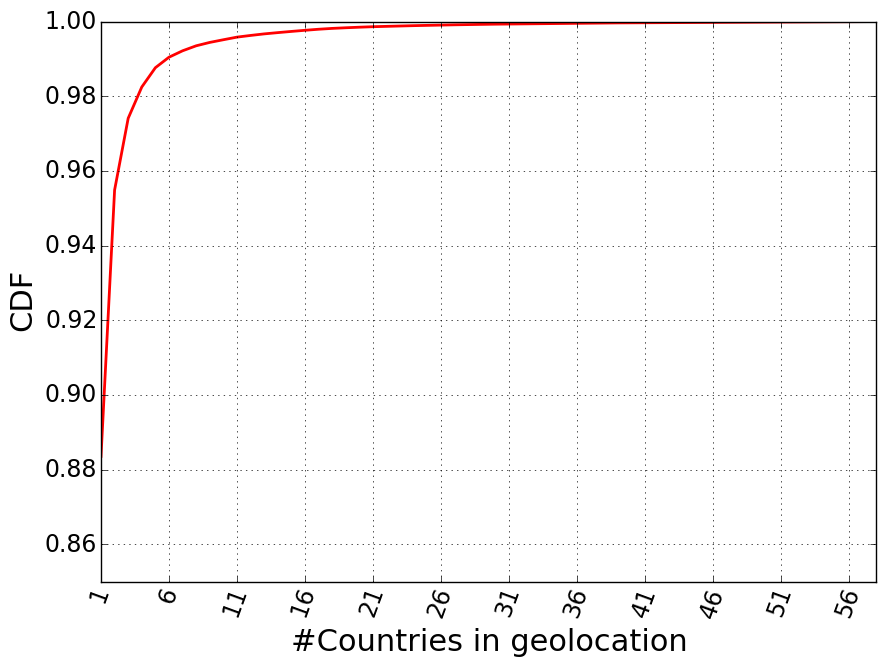}
	\caption{CDF: Number of countries in AS geolocation}
	\label{fig:CDF_ASNPrefixCountry-16}
\end{figure}

Finally, we map an AS to a set of countries by taking a union of all the 3 steps above. This is the merge step in Figure~\ref{fig:asgeoflowchart}. The distribution of AS geolocation is shown in Figure~\ref{fig:CDF_ASNPrefixCountry-16}. Perhaps surprisingly, only about 11.6\% ASes out of a total of 52,984 geolocated to multiple countries. We believe that this is the result of a practice where most organizations use a different AS number in different countries. If an AS does geolocate to multiple countries we use the set of all countries in our analysis. We could not geolocate 24 ASes because none of their BGP prefixes could be geolocated, no infrastructure IP from our set mapped to it nor did we find its IXP presence in public datasets. These ASes on an average announced only 2 to 3 BGP prefixes.\\

\begin{table*}[htb]
\centering
	\caption{Comparison of CAIDA's AS Rank with number of countries in AS Geolocation}
\begin{tabular}{C{1.5cm}: C{1cm}: C{3.5cm}: C{6cm} :C{2cm}}
		\hline
AS Rank & ASN & Customer Cone Size  & AS Name  & \#Countries\\ \hline
1 & 3356 & 24,553  & Level 3 Communications  & 63   \\
2 & 174 & 17,891 & Cogent Communications & 58   \\
3 & 3257 & 16,963 & Tinet Spa & 34 \\\hdashline
998 & 25394 & 18 & MK Netzdienste GmbH Co. KG & 2 \\ 
999 & 6724 & 18 & Strato AG & 4 \\ 
1000 & 52925 & 18 & Ascenty DataCenters Locacao e Servicos LTDA & 2 \\ 
\hline
		\end{tabular}
	\label{table:asrank}
\end{table*}

\noindent
\textbf{Comparison with CAIDA's AS Rank:}\\
Although our end goal is to detect detours, these geolocation results provide interesting insights. To understand more about which ASes geolocate to more than one country we use CAIDA's AS Rank~\cite{asrank}. This dataset gives higher ranks to ASes that have large customer cones. Intuitively, ASes with higher rank should resolve to many countries due to their wider presence. Table~\ref{table:asrank} shows ASes with their CAIDA AS rank and corresponding number of countries the AS geolocated to for top 3 and bottom 3 in the first 1000 ranked ASes. As expected, we see that ASes which have large presence with many customers across the world geolocate to large number of countries and low rank ASes with smaller customer cones geolocate to fewer countries.

%\subsection{AS to Country Validation}
%TODO: Create a ground truth for AS geolocation and give percentage of how many we correctly geolocate.
\squeezeup{1}
% !TEX root = main.tex
\section{Detour Detection}
\label{sec:pathanalysis}
We define a path as having a detour if the origin and destination is country \emph{`A'} but the path unambiguously includes some other country \emph{`B'}. Note that this approach examines paths where the prefix origin AS and the AS where the peer is located are in the same country. To analyze the AS path, we provide the following definitions:
\begin{itemize}
	\item \textbf{Prefix Origin}: The AS that announces the BGP prefix.
	\item \textbf{Detour Origin AS}: The AS that starts a detour in country \emph{`A'} and diverts the path to foreign country \emph{`B'}.
	\item \textbf{Detour Origin Country}: The country where we approximate location of Detour Origin AS, country \emph{`A'}.
	\item \textbf{Detour Destination AS}: The AS in foreign country \emph{`B'}.
	\item \textbf{Detour Return AS}: The AS where detour returns back in country \emph{`A'}.
\end{itemize}

%Detour path example
\begin{figure}[htb]
\centering
%\begin{minipage}[b]{.43\linewidth}
	\includegraphics[scale=0.6]{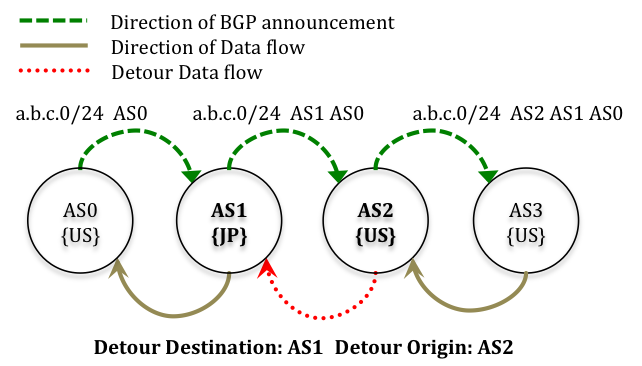}
	\caption{Example showing direction of BGP announcement and direction of observed detour}
	\label{fig:detourpathexample}
\end{figure}

Figure~\ref{fig:detourpathexample} illustrates detours. \textit{AS0} announces prefix \textit{a.b.c.0/24} to AS1, AS2 and AS3. AS1 geolocates to JP whereas AS3, AS2 and AS0 are in the US. In this case, data traversing from AS3 to AS0 will contain a detour from AS2 (Detour Origin) to AS1 (Detour Destination). We do not include sub-paths in our analysis; other portions of the path that may experience a detour. For example, in path AS1\{US\}-AS2\{IN\}-AS3\{CN\}-AS4\{IN\}-AS5\{US\}, we only count the detour US-IN-US. We do not count the detour IN-CN-IN.

There are some cases where we need to approximate detour origin and country. In a path such as AS1\{US\}-AS2\{US,BR\}-AS3\{CN\}-AS4\{US\}. We resolve the uncertainty of the detour origin by assuming that it starts in AS2, since there is a likely path to AS2 from AS1 through the US and AS2 starts the detour from US, not BR. We do not characterize \textbf{\textit{possible}} detours. For example, a path that geolocates to \{US\}-\{US,IN\}-\{US\} may in fact stay within the US and never visit India. In this work we only focus on paths that contain \textbf{\textit{definite}} detours, such as \{US\}-\{IN\}-\{US\} or \{US\}-\{IN,CN\}-\{US\}. Again, we re-emphasize that in this work we only look at paths that confidently start and end in the same country; paths like \{US,BR\}-\{IN\}-\{US\} or \{US\}-\{IN\}-\{BR\} are not considered. 
%No geolocation ASes
We discard paths where we see an AS whose geolocation is unknown and a detour is not certain. For example, paths like AS1\{US\}-AS2\{\}-AS3\{US\} are discarded. However, if we see the detour occurring before the AS that could not be geolocated we do count it as a valid detour i.e., in AS1\{US\}-AS2\{BR\}-AS3\{US\}-AS4\{\}-AS5\{US\}, AS4 does not have geolocation information but the US-BR-US detour occurred earlier. We treat this path as definite detour. 
%Country codes and how we use them
We note that in addition to geolocation accuracy there is also some ambiguity about exact country boundaries. Some territories and relationships are currently disputed between multiple authorities and no worldwide consensus exists. For example, Hong-Kong and the People's Republic of China could be considered one or two entities. Hong-Kong is affiliated with China but it is a charter city and has its own independent constitution and judiciary system. For our analysis, we left the resolution of boundaries and countries to the MaxMind database. With this particular example, Hong-Kong and China are treated as two separate entities. MaxMind follows ISO 3166 country codes. In some cases the geolocation from MaxMind is ambiguous: `A1:Anonymous Proxy', `A2: Satellite Provider', `O1: Other Country', `EU: Europe', `AP: Asia/Pacific'. We discard detours caused by these ambiguous codes, such as \{DE\}-\{EU\}-\{DE\}.\\

\noindent
\textbf{Filtering peered AS paths:}\\
%\label{sec:quantifyingpeeringrelations}
It is possible that the detour origin and the detour return ASes have a peering relationship and in reality traffic was not detoured at all. This, however, is hard to determine with certainty since peering relations and policies are not public. What we can do is provide an upper bound on how many detours may be eliminated due to peering. To detect such cases we use CAIDA's AS relationship dataset~\cite{caida:asrel}. This dataset provides information of provider to provider (p2p) and provider to customer (p2c) relationship between ASes. We count cases where p2p link might be used, i.e., data originates from the peer itself or from a downstream customer. In case of p2c link we assume this link is always chosen. We eliminate such paths from our analysis and revisit this issue in the next section summarizing the peering relationships in Table \ref{table:peeringinfo}.\\

%%Peering Example
%\begin{figure}[htb]
%\centering
%	\includegraphics[scale=0.6]{figures-2014/PeeringExample.png}
%	\caption{Example showing peering of ASes and RouteViews peer}
%	\label{fig:peeringexample}
%\end{figure}
%\squeezeup{2}

\noindent
\textbf{Multi-Origin Prefixes:}\\
Some prefixes are announced by more than one ASes. We do not eliminate such cases. So, if a prefix \emph{a.b.c.0/24} is seen in RIBs of 2 peers with AS paths `X Y Z' and `P Q R' then we treat each path as independent and detect detour if it fits above mentioned criteria of starting and ending in the same country. In our geolocation dataset we observed 7,579 prefixes of multi-origin (7,247 originated from 2 ASes). Out of these 6,104 suffered a detour. Motivation to not eliminate these prefixes is as follows: Network operators of such prefixes might want to re-evaluate their decisions especially if the ASes originating the prefix are in different countries. This might be a cause of high latency. 

%TODO: How to account for AS path poisoning?
\squeezeup{1}
% !TEX root = main.tex
%Talk about using the ground truth datasets
\section{Detour Validation}
\label{sec:detourvalidation}
In this section we validate detours in near real time using \emph{traceroutes} from RIPE Atlas probes. Our validation comprises of four steps:
\begin{enumerate}
\item{Run \texttt{Netra} with live BGP feeds from 416 peers to detect detours.}
\item{When a detour is detected, run corresponding traceroutes (from same country and same AS) using RIPE Atlas.}
\item{Check if the traceroute and detour see similar AS path.}
\item{Validate using traceroute IP hops and RTT.}
\end{enumerate}
\begin{figure*}
\centering
	\includegraphics[scale=0.5]{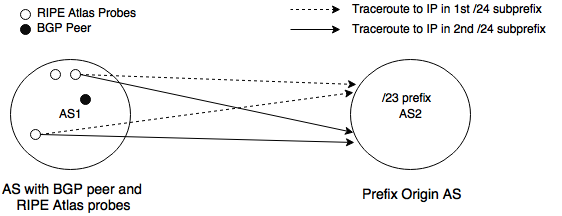}
	\caption{Data plane measurements: Example showing selection of RIPE Atlas probes and target IPs}
	\label{fig:traceroutemethod}
\end{figure*}
\squeezeup{1}
\subsection{Data-plane Measurements}
\label{sec:dataplanemeasurements}
We ran \texttt{Netra} from May $2^{nd}$ 2016 noon to midnight (using BGP feeds from 416 peers). When a detour was detected in control plane we selected RIPE Atlas probes in the same country and same AS which we detected detour from and ran traceroute (ICMP Paris-traceroute\cite{icmp-paris}) to IP addresses in the detoured prefix. The methodology to run data plane measurements is shown in Figure~\ref{fig:traceroutemethod}. There are a few cases where more than two Atlas probes are present in selected AS; in this case we selected 2 probes that are geographically farthest from each other. By doing this we aimed to account for cases where routes seen from geographically distant vantage points within the same AS are different. To select target IPs from detoured prefix, we break the prefix into its constituent /24s and randomly select an IP from each /24. For example, in a /23 prefix we select 2 IPs belonging to different /24s. By doing this we account for cases where a large prefix, even though in the same country, has different connectivity via different upstream provider. During this live run we detected 6,175 detours. Out of these 5,787 were unique detours (\{peer,prefix,aspath\} tuple).
\squeezeup{1}
\subsection{Selecting Congruent Paths}
Only 72 peers saw the 6,175 detours and the 72 peers belong to only 63 ASes. From these 63 ASes we then select ASes that also have active RIPE Atlas probes; there were only 10 ASes that both saw a detour and host a RIPE Atlas probe. 169 detours were seen from these 10 ASes corresponding to 6 countries: \{Brazil, Italy, Norway, Russia, United States, South Africa\}. From the 169 traceroutes we initiated to detoured prefixes, we discard 6 traceroutes where less than 3 hops responded since drawing detour conclusion from these is not possible. Finally, we are left with 163 traceroutes that can be used for validation. We acknowledge that 163 is not a very large number for validation purposes. However, running \texttt{Netra} for more hours does not necessarily increase the number of usable traceroutes for validation by a lot, we are limited by the number of ASes that have RIPE Atlas probes which also see a detour and detour-origin and detour-destination have no peering.
%The reason we chose to use only 12 hours of live data is that 

In total we detected 85 prefixes (corresponding to the 163 traceroutes) that suffered a detour that was visible from an AS which has RIPE Atlas probes. Note that some detoured prefixes were larger than /24, so we traceroute multiple IPs within it as explained in Section~\ref{sec:dataplanemeasurements}. The validation methodology is stated in Algorithm~\ref{algo:netraAlgo}. As previous work~\cite{incongruities} has pointed out, we found many cases where AS path seen in control plane and AS path seen in data plane do not match. However, these paths can still show detour if the detour origin AS and the detour destination AS are still present in the traceroute observed AS path. We call such AS paths \emph{congruent}. More specifically, we consider the detoured AS path congruent only if detour origin AS and detour return AS both are present in the traceroute-observed AS path in the same order (detour origin first). For example, if an AS path `\texttt{A B C D E}' in control plane changed to `\texttt{A X B C E}' in data plane where `\texttt{B}' was detour origin and `\texttt{C}' was detour destination, we consider it as a congruent path. To resolve traceroute path to AS path we used CAIDA ITDK and iPlane IP to AS mappings and in cases where no match was found we use longest prefix match on the global routing table for the hop IP. Then we map the longest prefix match to the AS that originated it. 
%As stated in Section~\ref{sec:prefixgeolocation} we believe an AS to own a prefix only if it announced it for more than 15 days in our one month dataset. While this is not an accurate method to determine ownership of a prefix, we believe most hijacks will be fixed within this window. The methodology to map traceroute to AS path is shown in Figure~\ref{fig:traceroutetoASpath}. 
Out of all the IPs we saw in 163 traceroutes, only 44 could be mapped to an AS using the IP to AS datasets. All other IPs were mapped using longest prefix match. 

We observed 113 congruent AS paths. This includes 3 cases, insertions, deletions and mix of both. We borrow nomenclature of these paths from~\cite{incongruities}. We saw 73 deletions, 29 insertions, 4 mix of insertion and deletions. The remaining 7 AS paths were exact matches. Note that these insertions and deletions occurred only for ASes that were not involved in the detour.
%Traceroute to AS Path
%\begin{figure*}[h]
%\centering
%	\includegraphics[scale=0.6]{figures-2016/traceroutetoASpath.png} 
%	\caption{Example showing mapping from traceroute to AS Path}
%	\label{fig:traceroutetoASpath}
%\end{figure*}

\begin{algorithm}
\caption{\texttt{Netra} Validation}\label{euclid}
\label{algo:netraAlgo}
\begin{algorithmic}[1]
\Procedure{validateASPath}{}
\State $\textit{aspath}\gets \textit{AS Path from Traceroute}$
\State $\textit{doas}\gets \textit{Detour Origin AS from \texttt{Netra}}$
\State $\textit{ddas}\gets \textit{Detour Destination AS from \texttt{Netra}}$
\If {\textit{doas},\textit{ddas} in \textit{aspath}}
\If {\textit{doas} before \textit{ddas} in \textit{aspath}}
\State Return \textit{True}
\EndIf
\EndIf
\EndProcedure
\end{algorithmic}

\begin{algorithmic}[1]
\Procedure{validateIPHops}{}
\State $\textit{ipHops}\gets \textit{IP hops from Traceroute}$
\State $\textit{ipHopCountries}\gets \textit{MaxMind-paid}$
\If {\textit{ipHopCountries} show detour}
\State $\textit{detourDestTR}\gets \textit{Dest. from traceroute}$
\State $\textit{detourDestNetra}\gets \textit{Dest. from \texttt{Netra}}$
\If {\textit{detourDestTR} in \textit{detourDestNetra}}
\State Return \textit{True}
\EndIf
\EndIf
\EndProcedure
\end{algorithmic}

\begin{algorithmic}[1]
\Procedure{validateRTTs}{}
\State $\textit{hopRTTs}\gets \textit{RTTs from Traceroute}$
\If {\textit{hopRTTs} show magnitudeJump}
\State Return \textit{True}
\EndIf
\EndProcedure
\end{algorithmic}

\begin{algorithmic}[1]
\Procedure{main}{}
\BState\hspace{\algorithmicindent} \textit{loop}: Each Detected Detour
\If {validateASPath}
\State {validateIPHops}
\State {validateRTTs}
\EndIf
\EndProcedure
\end{algorithmic}
\end{algorithm}
\squeezeup{1}
\subsection{Validation}
\begin{figure}[h]
\centering
	\includegraphics[scale=0.6]{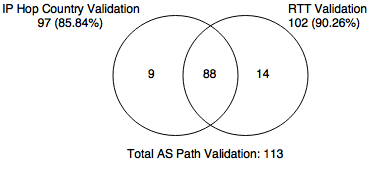}
	\caption{Validation Results: Live traceroutes using RIPE Atlas}
	\label{fig:validationresult}
\end{figure}

Now we validate detours detected by our methodology by comparing it with detours seen in data plane. For the 113 congruent AS paths, we evaluate if a data plane detour was seen. We chose to perform two tests. First, we resolve IPs observed in the hops of traceroute to country level geolocation using Maxmind-paid. We detect data plane detour if a path traversed foreign country and returned. We make sure that country visited (detour destination country) in data plane is present in the set of destination countries expected for this particular detour by \texttt{Netra}. We do this filtering to avoid false positives like: \texttt{Netra} detected detour \{US\}--\{GB,DE\}--\{US\} and traceroute detected detour \{US\}--\{IT\}--\{US\}. Although still a detour, since it was not accurately captured we count it as a miss. However, no such case was found. 
Second, we validate using RTT measurements. We detect RTT based detour if a hop in the traceroute showed increase in RTT by an order of magnitude (at least 10 times increase). The results of this analysis are shown in Figure~\ref{fig:validationresult}. We observed accuracy of about 85\% (97 out of 113) in country-wise method and 90\% (102 out of 113) by RTT measurements. The overlap between these two different tests was also large. 88 detours were detected in both (77.8\%). 

We investigate further the 9 detours that were seen in country-wise method but not in RTT. These detours covered small geographic area; 4 from Italy to France, 2 Norway to Sweden, 2 from Brazil to US and 1 from Russia to Sweden. RTTs between these countries have been previously reported to be low. Next we investigate 14 cases which were captured in RTT measurements but not in country-wise method. All of these do cross international boundaries. For 12 of these cases, due to large number of traceroute hops (especially towards the end of the traceroute) not responding we don't see the route returning to the origin country, hence not detected by country-wise method. We attribute remaining 2 cases as false positives due to inaccurate AS geolocation. 

%Validation Venn
\begin{figure*}
\centering
	\includegraphics[scale=0.4]{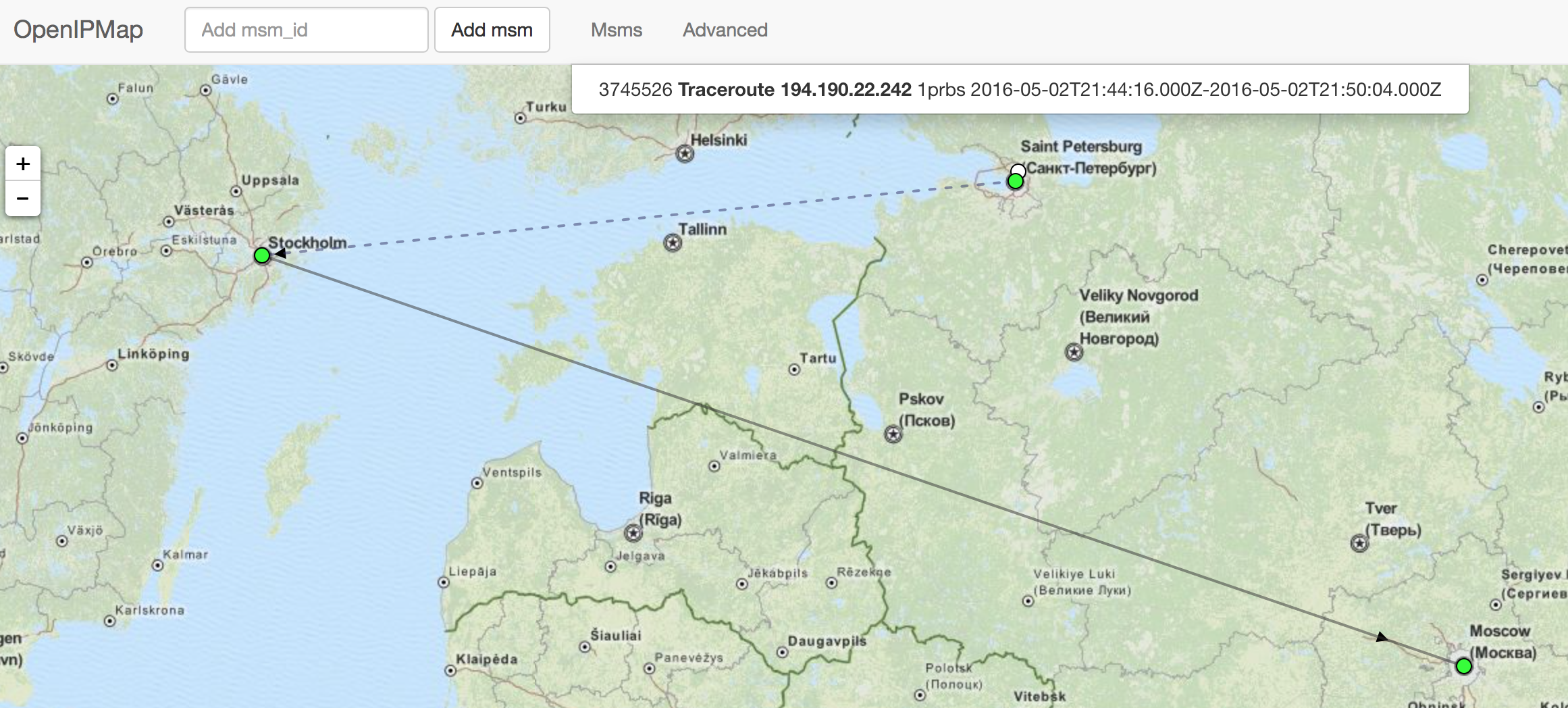}
	\caption{Top Detour on May $2^{nd}$ 2016: Detected using \texttt{Netra}, visualization using \texttt{OpenIPMap}. Dotted arrow represents multiple hops and solid arrow represents direct hop.}
	\label{fig:detourviz}
\end{figure*}

%Example Detour
In Figure~\ref{fig:detourviz} we provide a visualization of the most common detour we observed from Russia. Only visualization is done using OpenIPMap.\\

\noindent
\textbf{Validation Discussion:}\\
We show that large percentage of detours seen in control plane are accurately reflected in data plane as well. The main challenge is AS paths in both data plane and control plane don't agree in about 30\% cases. We note that this could be an artifact of Atlas probes connected differently than the peers which provide BGP feeds. It is, however, possible to learn common AS insertions and deletions over a period of time and evolve detection capabilities. 
%We revisit this future work in Section~\ref{sec:conclusion}.
\squeezeup{1}
\section{Results}
\label{sec:results}
In this section we quantify detours detected in January 2016. First, in Section~\ref{sec:aggresults} we present an overview of all the detours detected in our dataset. In Section~\ref{sec:detoursclassification} we define metrics and classify detours based on their stability and availability. In Section~\ref{sec:characterizingtransientdetours} we focus on transient detours.
\squeezeup{1}
%!TEX root = main.tex
\subsection{Aggregate Results}
\label{sec:aggresults}
\begin{figure}[h]
\centering
  \includegraphics[width=2.5in,height=1.6in]{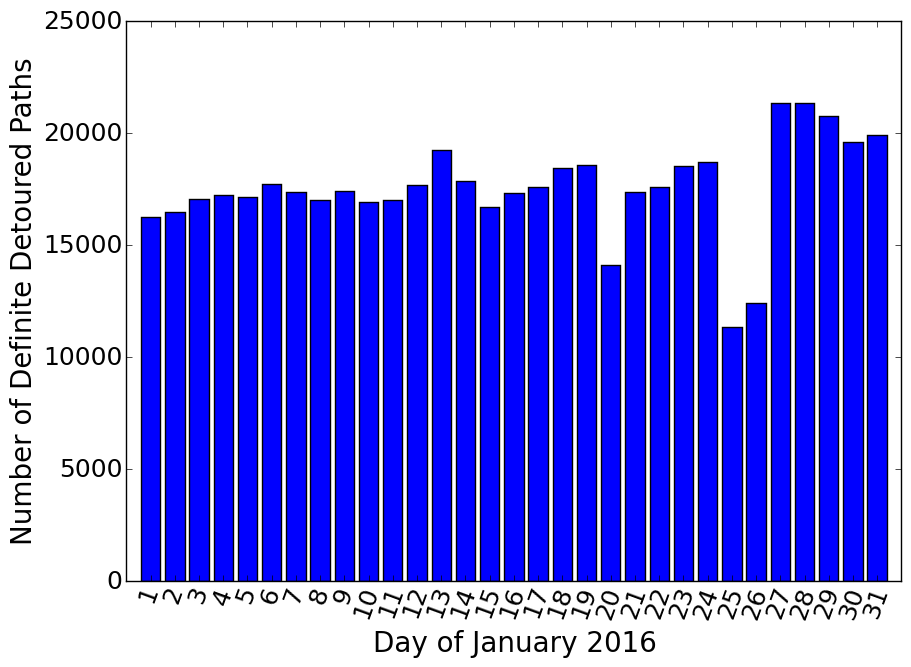}
	\caption{Total number of definite detours per day in January 2016}
	\label{fig:defAbnormal-16}
\end{figure}

%Talk about some aggregate numbers
We begin by characterizing aggregate results, namely all detours seen by all peers; in other words, we count an incident every time an AS path appears in a RIB of any peer that contains a detour. Many of these incidents are duplicates. Therefore in addition to the total we also present the number of unique detours. As expected, we observe that detours are not generally common. Also, not all peers see a detour. Only 79 peers, out of 416, saw one or more detours. Table~\ref{table:aggstats} details the number of detours seen. We analyzed about 14 billion RIB entries and about 544K entries showed a detour; out of theses only 18.9K were unique (most detours re-appear during the month). Figure~\ref{fig:defAbnormal-16} shows the number of detours for each day in January 2016. On an average we find about 17.5K detoured entries per day. 

\begin{table}[h]
\colspace{1.8}
\centering	
	\caption{Aggregate number of detours detected}
\begin{tabular}{C{3cm}:C{1.8cm}:C{1.5cm}}
\hline
\#Total RIB Entries & \#Detoured Entries & \#Unique Detours\\ \hline
14,366,653,046 & 544,484 & 18,995\\ \hline
		\end{tabular}
	\label{table:aggstats}
\end{table}

\begin{table}[h]
\colspace{2}
\centering	
%\small
\caption{Routes that may have peering relations}
\begin{tabular}{C{3cm}:C{2.5cm}:C{1.1cm}}
		\hline
\#Total Detours without filtering peered paths & \#Detours with possible peering & \%\\ \hline
659,569 & 115,085 & 17.4\%  \\\hline
\end{tabular}
	\label{table:peeringinfo}
\end{table}

%Talk about number of ASes that cause most detours and prefixes
Next we examine the visibility of detours, where we observe an uneven distribution among ASes. Just 9 ASes originate more than 50\% of the detours. Similarly, some prefixes experience detours more than others. 132 prefixes experienced more than 50\% of the total detours. Looking at the average length of a detour, we see that a detour visits 1 to 2 foreign ASes before returning to its origin country. \\
%The maximum length of a detour we observed was 4 - 5 hops. We do not characterize the geographical distance of detours, a task not supported by our data.

%Peering Info
\noindent
\textbf{Impact of Peering:}\\
We now estimate the effect of peering links on detours. Specifically, we are interested in cases where a peering relationship exists between the \emph{Detour Origin AS} and the \emph{Detour Return AS} as described in Section \ref{sec:pathanalysis} using CAIDA AS relationship dataset. If such a link exists, it is possible that traffic traverses that link instead of the detour. Table~\ref{table:peeringinfo} shows the number of detours between ASes that also have peering relations compared to total number of detours without filtering peered paths. We find that 17.4\% of the detours are avoided due to peering relations. We do not count these as detours in our analysis.\\

%Top Detour Orgins and Prefixes 
\noindent
\textbf{Top Detour Origins and Prefixes:}\\
To understand more about the nature of these detours, we focus on the origin and destination ASes. In Table~\ref{table:asnResponsible} we show the common detour origins and country where the AS was approximated to origin the detour from. Next is the percentage of detours out of the total that started from given origin. Following the percentage, is the most frequent destination that was visited from the origin, and lastly is the percentage of detours that went to most common destination from the said origin. We observe that most commonly these were access provider ASes.
Similarly, in Table~\ref{table:detouredPrefix} we show top impacted prefixes.\\
%TODO: Talk more about top affected prefixes.

\begin{table*}
\colspace{2}
\centering	
	\caption{Top Detour Origin ASNs for all detoured paths}
\small
\begin{tabular}{C{3.3cm}:C{1.2cm}:C{3.5cm}:C{2cm}}
		\hline
Top Detour Origin AS & Total \% & Frequent Detour Destination AS & \% to frequent destination\\ \hline
3356 \captionnote(Level 3 Communications,BR) &8.39\% &32787 \captionnote(Prolexic-Technologies DDoS Mitigation Network) &30.99\%  \\\cdashline{1-4}
12956 \captionnote(Telefonica International Wholesale Services,BR) & 5.74\% & 262182 \captionnote{Media Networks Latin America} & 46.33\%  \\\cdashline{1-4}
6939 \captionnote(Hurricane Electric,US) & 4.99\% & 45932 \captionnote(Net Sys International Limited) & 15.9\%  \\\cdashline  {1-4} \hline
		\end{tabular}
	\label{table:asnResponsible}
\end{table*}

\begin{table*}
\colspace{2}
\centering	
	\caption{Top Detoured prefixes and corresponding percentages}
\small
\begin{tabular}{C{3.3cm}:C{1.2cm}:C{3.5cm}:C{2cm}}
%\begin{tabular}{C{2cm}:C{3cm}:C{1cm}:C{1.8cm}:C{1cm}}
%\begin{tabular}{ m{0.8cm} m{1.1cm} m{0.7cm} m{0.7cm} m{2cm} m{0.7cm} }
		\hline
Prefix Affected & Total \% & Frequent Detour Destination AS & \% to frequent destination\\ \hline
199.253.181.0/24 \captionnote(Internet Systems Consortium,US) & 0.51\% & 766 \captionnote(Entidad Publica Empresarial Red) & 100\%  \\\cdashline{1-4}
167.220.28.0/23 \captionnote(Microsoft,US) & 0.51\% & 6584 \captionnote(Microsoft Corp) & 100\%  \\\cdashline{1-4}
199.6.5.0/24 \captionnote(Internet Systems Consortium,US) & 0.51\% & 766 \captionnote(Entidad Publica Empresarial Red) & 77.11\%  \\ \hline
\end{tabular}
	\label{table:detouredPrefix}
\end{table*}

\begin{figure}
\centering
	\includegraphics[width=2.5in,height=1.6in]{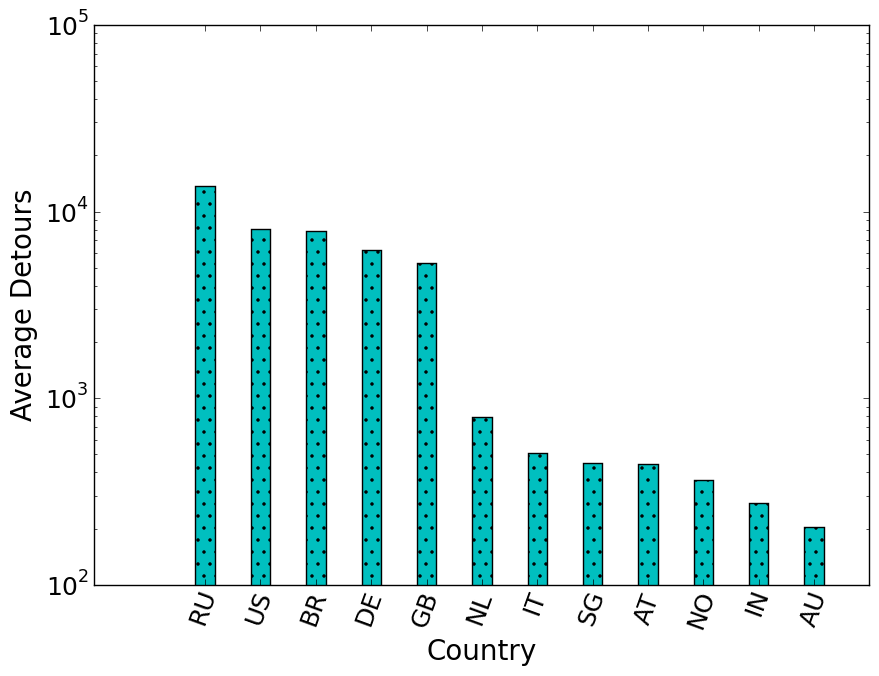}
	\caption{Average number of detours per country}
	\label{fig:histAllDetourspercountry}
%\end{minipage}
\end{figure}

\noindent	
\textbf{Country-wise analysis:}\\
To provide an understanding on number of detours per peer in each country we normalize the data by dividing the number of detours by number of peers in the country. The reason to normalize data is simple, RouteViews and RIPE RIS peers are not evenly distributed among different countries. Therefore it is possible that more detours are seen in countries that have more peers due to more visibility. An average number of detours per peer per country provides better insight. Out of 30 countries, only 12 countries observed a detour. Figure~\ref{fig:histAllDetourspercountry} shows average number of detours per country. Russia showed most number of average detours. Understanding the total number of detours in different countries is important but it does not reflect if detours seen in different countries have different characteristics. In the next section we focus on characterizing these detours. 

\squeezeup{1}
% !TEX root = main.tex
\subsection{Characterizing Detours}
\label{sec:detoursclassification}
To characterize detours we define two metrics:

\begin{enumerate}
\item \textbf{Detour Dynamics}
	\begin{enumerate}
	\item \textbf{Flap Rate}: Measure of \emph{stability} of a detour; how many times a detour disappeared and reappeared.
	\item \textbf{Duty Cycle}: Measure of \emph{uptime} of a detour throughout the month measurement period.
	\end{enumerate}
\item \textbf{Persistence}: Total number of continuous hours a prefix was seen detoured.
\end{enumerate}

%Talk about peer filtering. One peer per AS.
Before using the above metrics to characterize the detours, we perform data pruning to avoid skewing of data towards ASes that have more peers that provide BGP feeds to RouteViews and RIPE RIS. Also, ASes with multiple peers and similar views can contribute duplicate detours to our dataset. We follow a simple approach to deal with this problem: if an AS contains more than one peer we select the peer that saw the most detours as the representative of that AS. This may potentially undercount detours since some peers in same AS may see different detours. After selecting a representative we are left with 36 (out of 79) peers. We now continue our characterization of detours by looking at \textbf{detour dynamics}. Specifically we focus on flap rate and duty cycle, defined as follows:
\squeezeup{3}
\begin{center}
$$\hspace{0.8cm}FlapRate=\frac{Total Transitions}{TotalTime}\times100$$
\end{center}
\begin{center}
$$\hspace{0.25cm}DutyCycle=\frac{Total Uptime}{TotalTime}\times100$$
\end{center}
These metrics provide insights into the life cycle of detours by measuring route uptime and stability. BGP route flapping is a known problem and has been studied in \cite{flapping1} by looking at BGP updates and RFC 2439 provides methods to dampen these. However, in context of this paper duty cycle and flap rate are calculated from the RIBs. We extract detours from the RIBs and evaluate when they disappear and reappear.

\begin{figure}[h]
\centering
	\includegraphics[width=2.5in,height=1.6in]{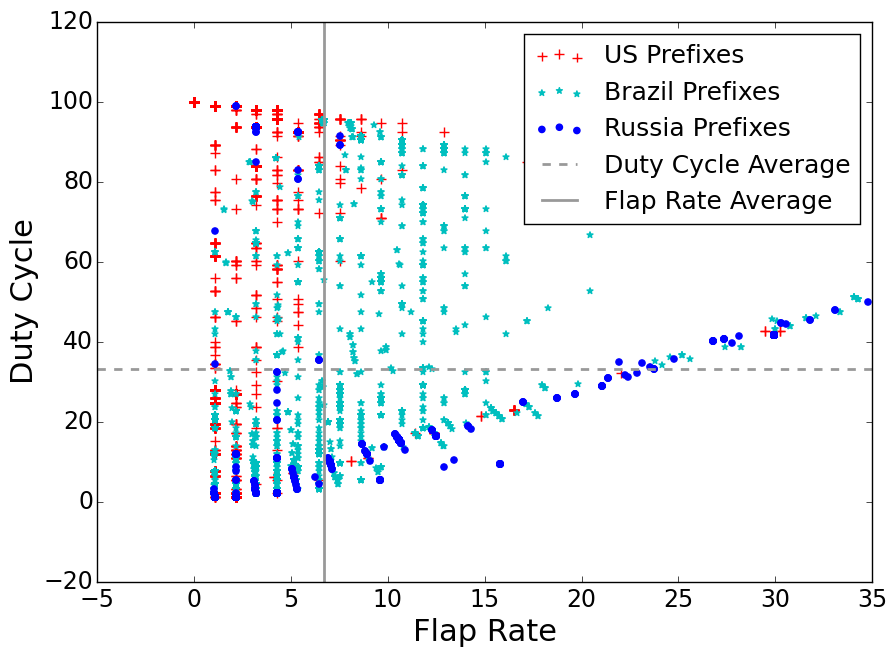}
	\caption{Flap Rate vs DC for US, RU and BR prefixes}
	\label{fig:dynamics16}
\end{figure}

\begin{figure}[h]
\centering
	\includegraphics[width=2.5in,height=1.6in]{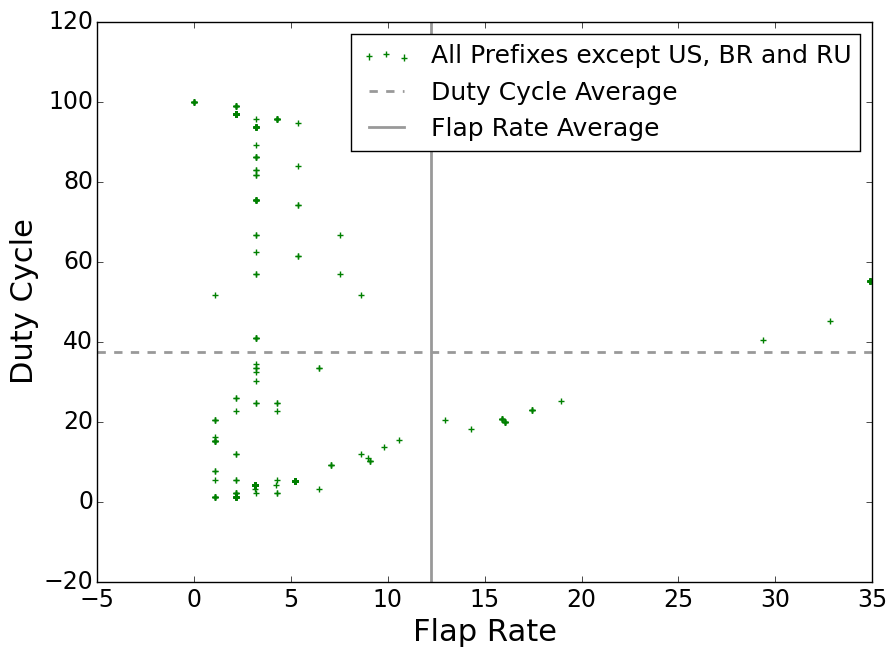}
	\caption{Flap Rate vs DC for Non US, RU and BR prefixes}
	\label{fig:dynamicsnon16}
\end{figure}

To understand if country where detours occur plays a role in detour dynamics, next we drill into country specific detours. Figure~\ref{fig:dynamics16} shows a scatter plot of flap rate vs. duty cycle for various detours in US, Brazil and Russia. We selected these three countries because they show the most detours in our dataset; they account for 93\% of detours. We see a triangular pattern with some outliers. Large number of detours show high duty cycle and low flap rate. We divide each figure into 4 quadrants based on average flap rate and average duty cycle of all detours. We name quadrants anti-clockwise starting from top right. US detoured paths appear more stable (lower flap rate and higher duty cycle) in $II^{nd}$ quadrant. On the other hand, Russian and Brazilian detoured paths fall mostly in the $I^{st}$, $III^{rd}$ and $IV^{th}$ quadrant. Russian detours in general showed lower duty cycle than US and Brazil. We also present a similar scatter plot for all the non US, BR and RU detours in Figure~\ref{fig:dynamicsnon16}. In this case we observed detours mostly in extreme ends on $II^{nd}$ and $III^{rd}$ quadrant indicating two categories of detours, either long lasting or very rare events. 
%Lastly, we looked at behavior of detours in South Africa and Kenya where lack of peering has been previously reported. Here, detours mostly lie in $II^{nd}$ quadrant indicating very stable detours (possibly intentional). 
A network operator can use information like this and decide which quadrant detours are more interesting to focus on. While all of detours may need attention, we believe detours with low duty cycle and low flap rate may need immediate attention. We talk more about this in Section~\ref{sec:characterizingtransientdetours}.

%DC and Flap Rate graphs

\begin{figure}
\centering
	\includegraphics[width=2.5in,height=1.6in]{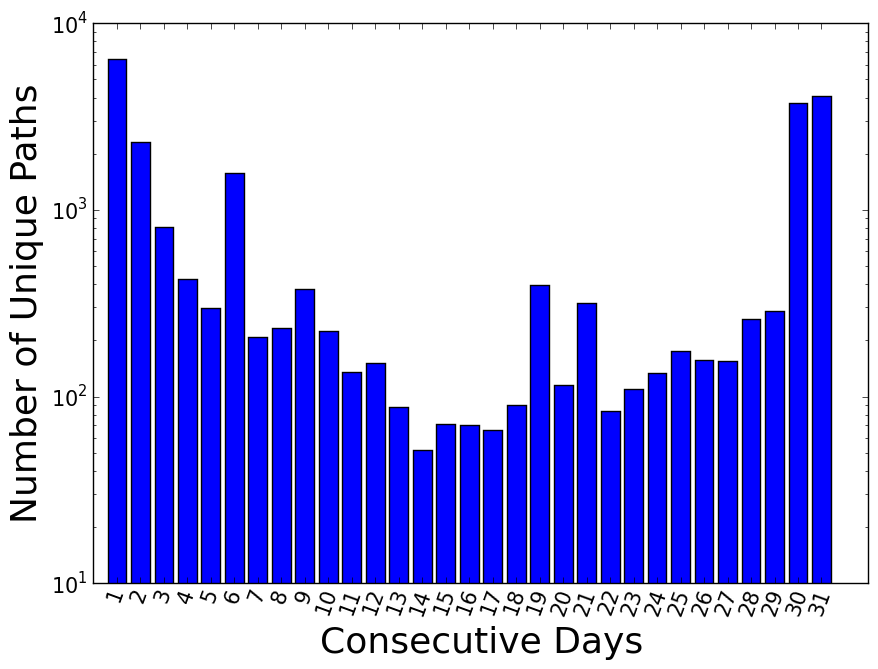}
	\caption{Persistence of definite detoured paths as seen by all peers}
	\label{fig:abnormalPersistenceAll-16}
\end{figure}

\begin{figure}[t]
\centering
	\includegraphics[width=2.5in,height=1.6in]{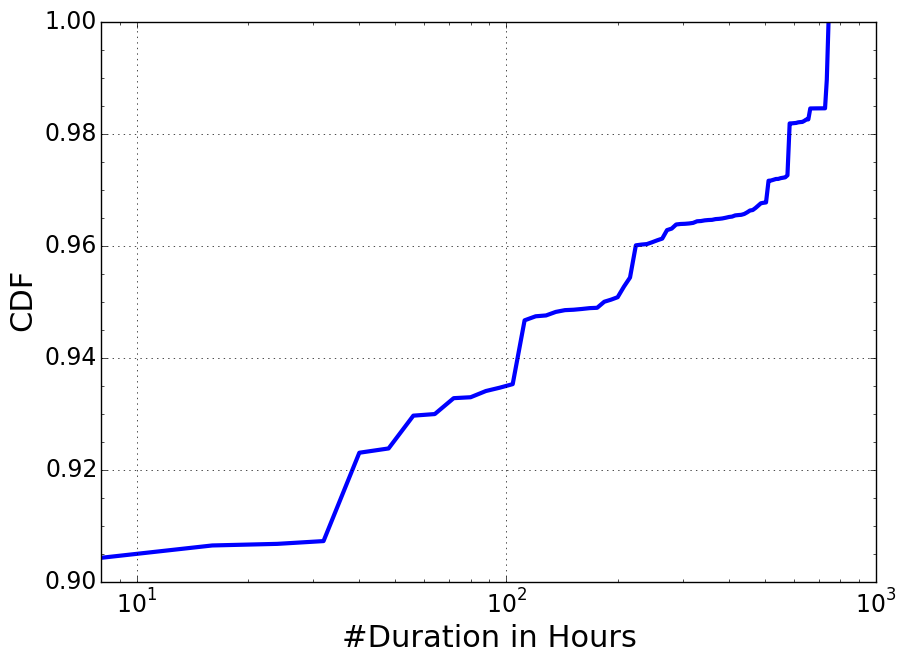}
	\caption{Distribution of detour duration}
	\label{fig:cdfdetoursduration}
\end{figure}

%Talk  about persistence metric.
Next, we examine the \textbf{persistence} of detours. Figure~\ref{fig:abnormalPersistenceAll-16} shows the number of consecutive days a detour was visible by any peer. Note that persistence is measured in number of consecutive hours hence captures different characteristics than duty cycle which measures uptime throughout the dataset. We see a U-shaped pattern in Figure~\ref{fig:abnormalPersistenceAll-16}, meaning that many detours are either short lived (one day) or they persist for entire month.  We take a different view at persistence in Figure~\ref{fig:cdfdetoursduration} by plotting CDF of duration in hours. We see that most detours are short-lived, with about 92\% lasting less than 72 hours, defined as \emph{transient} detours. 
%We justify the selection of 72-hour threshold as follows: this is the time between a misconfiguration, which is accidentally made on a Friday, and when it is fixed on Monday morning. We do realize that most networks will fix problems sooner than that, but there is a wide range in how networks are managed. 
Finally, we examine a specific case of a transient detour, namely \emph{flash detours} which appeared only once and never appeared again during the month.

In the following section we focus on transient and flash detours. Due to space limitations we do not characterize persistent detours further. We do note, however, that characterizing persistent detours is important for at least some of the reasons we enumerated earlier. We chose to focus on transient detours as they shed light on misconfigurations or even malicious activities, both aspects of routing we understand less.

%!TEX root = main.tex
\subsection{Transient and Flash Detours}
\label{sec:characterizingtransientdetours}

\begin{table*}
\colspace{2}
\centering	
	\caption{Top Transient Detour Origin ASNs}
\small
\begin{tabular}{C{3.3cm}:C{1.2cm}:C{3.5cm}:C{2cm}}
		\hline
Transient Detour Origin AS & Total \% & Frequent Detour Destination AS & \% to frequent destination\\ \hline
9002 \captionnote(RETN-AS RETN Limited,RU) & 22.64\% & 2914 \captionnote(NTT America) & 99.07\%  \\\cdashline{1-4}
6939 \captionnote(Hurricane Electric,IT) & 10.94\% & 8551 \captionnote(Bezeq International) & 100\%  \\\cdashline{1-4}
1299 \captionnote(TELIANET,IT) & 10.87\% & 8708 \captionnote(RCS-RDS) & 100\%  \\\cdashline  {1-4} \hline
		\end{tabular}
	\label{table:asnRespTransient}
\end{table*}

\begin{figure}
\centering
	\includegraphics[width=2.5in,height=1.6in]{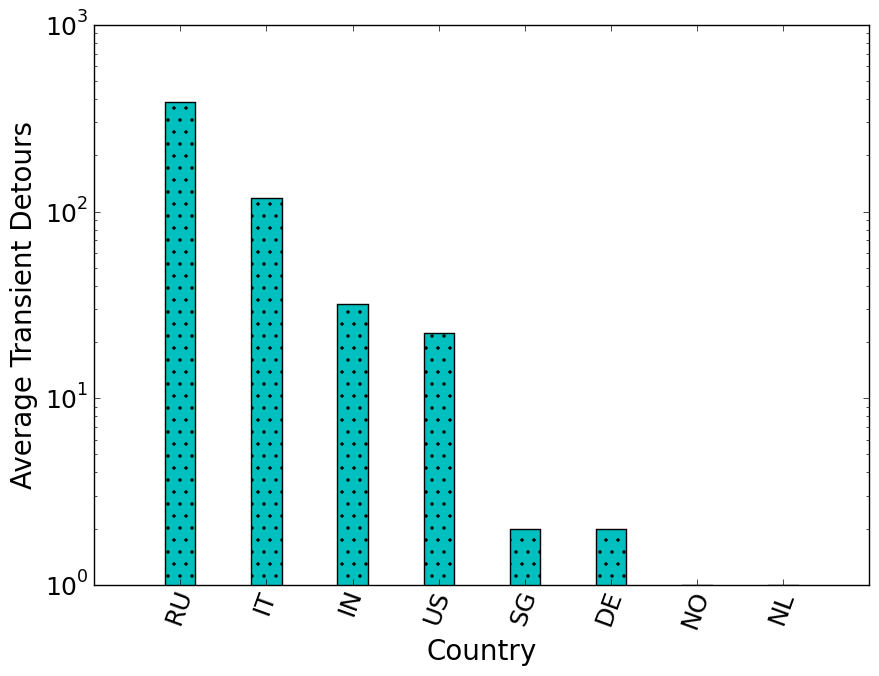}
	\caption{Average number of transient detours per country}
	\label{fig:transienthistpercountry}
\end{figure}

\begin{figure}[h]
\centering
	\includegraphics[width=2.5in,height=1.6in]{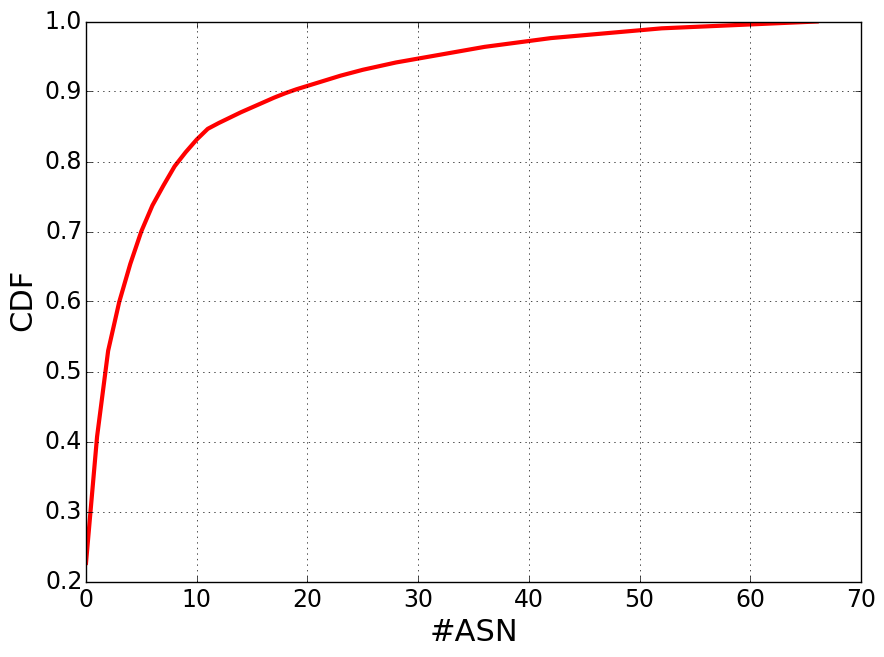}
	\caption{Distribution of ASes that originated a transient detour. The top 4 Detour Origin ASes account for 50\% of all transient detours}
	\label{fig:TrASNdetour_CDF}
\end{figure}

\begin{figure}
\centering
	\includegraphics[width=2.5in,height=1.6in]{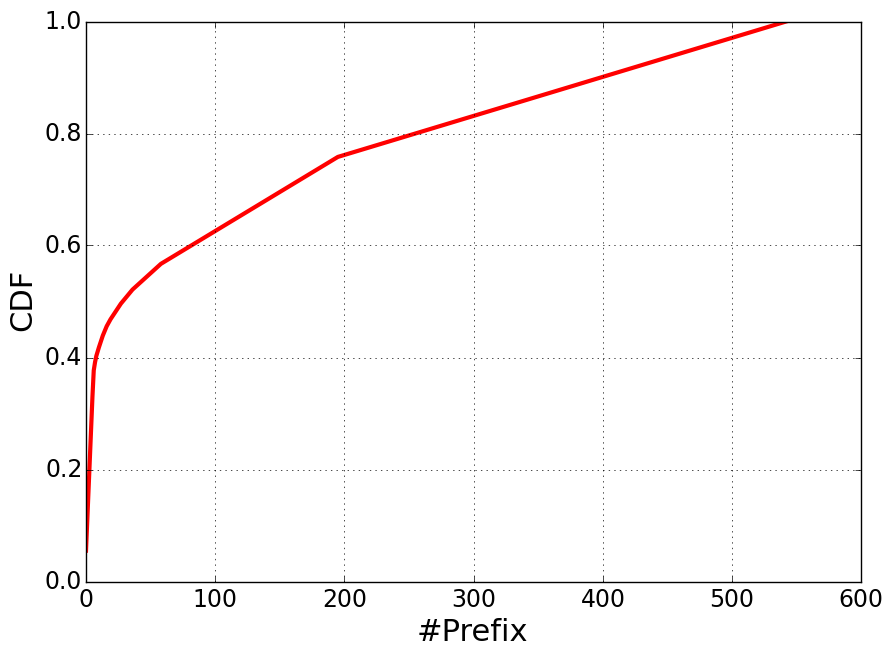}
	\caption{Distribution of prefixes that experienced a transient detour. About 30 prefixes account for 50\% of all transient detours}
	\label{fig:TrprefixCDF}
\end{figure}

We first present an understanding of the transient detours on per-country basis. Since there are more than one peers in some countries and different peers see varying number of transient detours, we calculate an average number of transient detours per country by dividing total number of transient detours in a country by number of peers in the given country. This average value per country is presented in Figure~\ref{fig:transienthistpercountry}. We detected transient detours in only 8 countries where Russia topped the list. In comparison to Figure~\ref{fig:histAllDetourspercountry} Italy and India showed more average number of transient detours than US. Figures~\ref{fig:TrASNdetour_CDF} and \ref{fig:TrprefixCDF} show a distribution of ASes that initiate detours and prefixes affected by detours. We observe that 4 ASes originate 50\% of the transient detours and only 30 prefixes account for  50\% of the transient detours.
Similar to Table~\ref{table:asnResponsible}, shown in Table~\ref{table:asnRespTransient} are the most common transient detour origins and Table~\ref{table:TransientprefixesAffectedTable} shows top impacted prefixes by transient detours. AS9002, RETN-AS, started the most number of transient detours in our dataset. We note that in \emph{ASWatch}~\cite{ASwatch} authors gathered ground-truth data from security blogs which enlisted AS9002 as a malicious AS. Another previously know malicious AS that appeared in our findings was AS49934 as a detour destination for 7 Russian prefixes. AS49934 is currently unassigned. It was assigned in Ukraine between 2009-10-14 and 2016-01-03 and was known to announce bogus prefixes and host bots. 

\begin{table*}
\colspace{2}
\centering	
	\caption{Prefixes affected the most by transient detoured BGP paths}
\small
\begin{tabular}{C{3.3cm}:C{1.2cm}:C{3.5cm}:C{2cm}}
		\hline
Prefix Affected & Total \% & Frequent Detour Destination AS & \% to frequent destination\\ \hline
178.79.218.0/23 \captionnote(Limelight Networks, Inc, IT) & 5.5\% & 8551 \captionnote(Bezeq International, IL) & 100\%  \\\cdashline{1-4}
185.19.164.0/22 \captionnote(Digi Italy S.R.L, IT) & 5.5\% & 8708 \captionnote(RCS-RDS, RO) & 100\%  \\\cdashline{1-4}
46.21.30.0/24 \captionnote(Tekka Digital, IT) & 5.5\% & 8758 \captionnote(Iway, CH) & 67.08\%  \\\cdashline{1-4} \hline
\end{tabular}
	\label{table:TransientprefixesAffectedTable}
\end{table*}

Finally, we look at \emph{flash} detours. These are detours that appeared only once and were observed in only one RIB of a peer. Flash detours account for 26\% of the transient detours, 328 prefixes (6\% of all prefixes that suffered detour) experienced at least one flash detour.
 
%Some interesting prefixes
Owners of the prefix which suffered flash detours might be interested to know such findings. While 328 prefixes suffered flash detours in our dataset, due to space limitation we point out a few interesting ones in Table~\ref{table:flashtable}.

\begin{table*}
\colspace{2}
\centering	
	\caption{Some prefixes affected by flash detours}
	\small
\begin{tabular}{C{3cm}:C{6cm}:C{3cm}}
%\begin{tabular}{ m{0.8cm} m{1.1cm} m{0.7cm} m{0.7cm} m{2cm} m{0.7cm} }
		\hline
Prefix Affected & Owner & Detour Destination\\ \hline
170.61.199.0/24 & Mellon Bank, US & 28513 \captionnote(Uninet, MX) \\ \cdashline{1-3}
192.230.0.0/20 & Washington State Department of Information Services, US & 7660\captionnote(Asia Pacific Advanced Network, JP)\\ \cdashline{1-3}
212.11.152.0/21 & Moscow Mayor Office, RU & 2603\captionnote(NORDUnet, NO)\\ \cdashline{1-3}
208.79.7.0/24 & Security Equipment Inc, US & 53185\captionnote(William Roberto Zago, BR)\\ \cdashline{1-3}
161.151.72.0/21 & The Prudential Insurance Company of America, US & 2510\captionnote(Infoweb Fujitsu, JP)\\ \hline
\end{tabular}
	\label{table:flashtable}
\end{table*}
%How can someone use the flash detour info?
The list in Table~\ref{table:flashtable} raises serious concerns. Data from government agencies, banks, insurance companies can easily be subject to wiretapping once it leaves national boundaries. Based on our control-plane only data, it is not possible to verify if these institutions were attacked or not. Nevertheless, we believe our findings will motivate network operators to look more closely into why their prefix detoured and if they intended it to happen.

\squeezeup{1}
% !TEX root = main.tex
\section{Discussion}
\label{sec:discussion}
In this paper we present a first attempt to characterize detours in the Internet. We sampled BGP routing tables from 416 peers around the world over the entire month of January 2016 to investigate international detours. We see about 18.9K distinct entries in RIBs that show a detour. More than 90\% of the detours last less than 72 hours. We also discover that a few ASes cause most of the detours and detours affect a small fraction of prefixes. Some detours appear only once. Our work is the first to present different types of detours, namely, persistent and transient. We also present novel insights on their characteristics such as detour dynamics in different countries, top impacted prefixes and detour origins.
%We find many detours to be persistent and most likely the result of traffic engineering decisions. We also find a significant portion, however, that are transient, appearing only for a few hours or days.

Characterizing detours in the Internet is very useful. Customers gain more insight into how their providers route traffic. There is perhaps an expectation from users that if they send traffic to other users in the same country the packets will not step outside national borders; our work provides evidence to the contrary. Network operators can use our methodology and results for diagnostic purposes. A sudden change in RTT may be traced to a detour, or keeping track of what the routing system does. The latter is important to assure customers that their traffic is not subject to monitoring by other governments.

Our work is useful to regulators and state officials responsible for network infrastructure, since our work quantifies information about a practice that may run afoul of state policy. State officials can use such information to assure citizens that their traffic stays within national borders or direct ISPs to alter their practices. State agencies that transmit sensitive information may monitor detours to alert for potential MITM attacks. For example, we did observe cases where prefixes belonging to US Washington state government were detoured through Japan; and for some detours from Russia malicious AS49934 appeared as detour destination.

Finally, entrepreneurs may use our results when deciding where to establish new Internet exchange points (IXP) or deploy infrastructure in developing countries.\\

%We have in our possession lot of information that network operators can leverage off. For example, one may want to know what the \emph{least} frequent detours are, or other one-off or rare events; or when a detour traveled a large geographic distance rather than a few AS hops; or interested in detours that last a very short time, or seen in certain parts of the world; network operators and users may want to know if their prefixes have experienced a detour as opposed to general statistics about all prefixes.
\noindent
\textbf{Dataset Contributions:}\newline
We make the geolocation and detours detection data available to the community via a public RESTful API interface. The motivation to do so is as follows. 1) Network operators can easily query our database and check if their prefix suffered a detour. 2) Internet measurement researchers can use this information to study various BGP anomalies such as route leaks, detecting malicious ASes, etc. Our results on AS and prefix geolocation are available at \url{http://geoinfo.bgpmon.io} and detours results can be accessed at \url{http://detours.bgpmon.io}.

% !TEX root = main.tex
\section{Conclusions and Future Work}
\label{sec:conclusion}
There is an increasing need, fueled by new national regulations in Europe and Australia, for ISPs to ensure that personal information belonging to their users does not leave the country. It is unclear whether such regulations cover data in transit as well as storage, but data can certainly be sniffed while in transit, violating the original intent. Such regulations may place a substantial burden on ISPs to prove that such data remains within a country for its entire lifetime, even when it moves. It is still far from clear what the implications are on ISP operations. Currently we do not have the tools to monitor data in transit and state with confidence that data has not left a country, even briefly.

Our work does not solve this problem. Rather, it lays the ground for an important conversation about the challenges new regulatory frameworks will pose to researchers, industry and network operators. Our work investigates only a small part of the problem, namely finding the subset of paths where we can detect international detours with some confidence. Our work provides some answers, but also brings attention to the problem and will hopefully stimulate more work in this new direction. The gauntlet was thrown and we expect a lot more research in this area.

Within its scope, we believe our work was executed carefully by taking into account measurements from both control and data planes. We show that for the cases were able to study there is agreement between the two planes. This is a significant result. Equally significant, our work has also illuminated the difficulties in expanding the scope within the existing measurement infrastructures. One of the main difficulties we encountered for example, is finding measurement points with both control (BGP peers) and data (RIPE probes) monitors to correlate results. This problem cannot be easily solved, it would take substantial effort to scale the existing infrastructures by an order of magnitude or more. Another important obstacle is lack of knowledge about peering relationships between ASes. This is also a hard problem to solve, since such relationships are not readily disclosed. It is interesting, however, to contemplate the issue if regulatory requirements require such disclosures.

Based on our results, we believe that it will be hard to solve this problem without substantial data plane monitor deployment to corroborate control plane measurements. ISPs and IXPs may be required to install sophisticated data plane probe infrastructures and geolocation databases may have to become far more accurate for infrastructure IP addresses in order to detect international detours with some certainty. Control plane monitoring is still very important as it provides efficient global monitoring and can immediately flag potential anomalies where data plane monitoring should be directed. Our work shows that it is effective and should be expanded.

In the future we plan to continue to build a system that detects international detours in real time. It is very apparent that we need to include both control and data plane measurements and study algorithms that take input from both. Our first goal is to provide ISPs with a tool to alert when a detour has taken place, followed by information about it (origin and destination AS, duration, source and amount of data in the ISP that followed the detour). We also plan to study emerging regulatory requirements and provide feedback about the challenges they pose.\\

%Our work raises interesting questions that span multiple research directions. Detected detours can be studied more to understand internet routing better. False positives (about 10\%) of our work could be a result of publicly unknown peering and IXP relationships on ASes. Mining these relationship will be useful to our work as well as other internet measurement researchers. We face the problem of control plane and data plane incongruities. In future, we plan to enhance our methodology to learn common cases of modifications to AS path in data plane and appropriately detect detours. With this paper we aim to fetch for participation from service providers to deploy our tool \texttt{Netra}, validate AS geolocation and detours to improve its detection capabilities. 

%Detecting detours in the Internet is an important but challenging problem. Raising alerts for such routing behaviors should be an integral part of BGP monitoring and defense systems. Public and regularly updated peering and IXP relationships, more BGP feeds to RouteViews and RIPE RIS and more RIPE Atlas probes are necessary for success of detour detection. Data plane measurements can compliment efforts in this direction but with control plane methodology performing large scale analysis becomes possible. \\

%\subsubsection*{Acknowledgement}
\noindent
\textbf{Acknowledgement}\\
We would like to thank Randy Bush (IIJ) for helpful discussion on AS geolocation; Emile Aben (RIPE NCC) for helpful insights on the paper and using OpenIPMap; Alessandro Improta and Luca Sani (Isolario Project, Italy) for providing code to crawl IXP websites; Roya Ensafi (Princeton) for help in geolocating infrastructure IP addresses.

\pushup{7}

\bibliographystyle{plain}
\bibliography{main.bib}

\end{document}